\documentclass[aps, prl, reprint,nofootinbib]{revtex4-2}

\usepackage{amsmath}
\usepackage{amssymb}
\usepackage{amsthm}
\usepackage{yfonts}
\usepackage{mathrsfs}
\usepackage{enumerate}
\usepackage{fullpage}
\usepackage{color}
\usepackage{verbatim}
\usepackage{graphicx}
\usepackage{dsfont}
\usepackage{hyperref}
\usepackage{tikz}
\usepackage{bm}
\usepackage{subfig}
\usetikzlibrary{arrows, positioning, shapes.geometric, decorations.markings, patterns}
\newcommand{\ket}[1]{| #1 \rangle}
\newcommand{\braket}[2]{\left\langle #1 \big\vert #2 \right\rangle}

\newcommand{\rank}{\operatorname{rank}}

\renewcommand{\tilde}{\widetilde}
\renewcommand{\bar}{\overline}
\newcommand\onenorm[1]{\lVert#1\rVert_1}

\newcommand{\tr}{\operatorname{Tr}}

\begin{document}

\title{Holographic Tensor Networks in Full AdS/CFT}
\author{Ning Bao}
\email{ningbao75@gmail.com}
\affiliation{Berkeley Center for Theoretical Physics, Berkeley, CA 94720}

\author{Geoffrey Penington}
\email{geoffp@stanford.edu}
\affiliation{Stanford Institute for Theoretical Physics, Stanford, CA 94305}

\author{Jonathan Sorce}
\email{jsorce@stanford.edu}
\affiliation{Stanford Institute for Theoretical Physics, Stanford, CA 94305}

\author{Aron C. Wall}
\email{aroncwall@gmail.com}
\affiliation{Stanford Institute for Theoretical Physics, Stanford, CA 94305}

\begin{abstract}
We present a general procedure for constructing tensor networks for geometric states in the Anti-de Sitter/Conformal Field Theory (AdS/CFT) correspondence. Given a state in a large-$N$ CFT with a static, semiclassical gravitational dual, our procedure produces a tensor network for the boundary state whose internal geometry matches (a discretization of) the bulk spacetime geometry. By invoking the ``holographic entanglement of purification'' conjecture, our construction can be made to capture the structure of the bulk spacetime at sub-AdS scales.
\end{abstract}

\maketitle

\section{Introduction}
In the AdS/CFT correspondence \cite{Maldacena1999}, it is well-known that information about the entanglement structure of certain large-$N$ CFT states can be computed by performing a gravitational calculation in a corresponding asymptotically-AdS spacetime \cite{RT2006-1, RT2006-2, HRT2007, Headrick2010, Dong2016, LM, DLR}. In particular, the von Neumann entropy $S = - \tr(\rho \log \rho)$ of the reduced density matrix $\rho$ for a subregion of the boundary state is given to leading order in $G_N$ by the area of the corresponding Ryu-Takayanagi (RT) surface \cite{RT2006-1}---or, more generally, the Hubeny-Rangamani-Takayanagi (HRT) surface \cite{HRT2007}---in the bulk spacetime.

It was proposed in \cite{Swingle2012-1, Swingle2012-2} that ``tensor networks'', discrete $D+1$-dimensional geometries used in condensed matter physics for the efficient classical simulation of $D$-dimensional quantum states, can be used as a toy model for understanding the holographic relationship between geometry and entanglement in AdS/CFT. In addition to universally satisfying a version of the Ryu-Takayanagi entropy formula \cite{Swingle2012-1}, tensor networks with AdS-like geometries have been shown to display features of quantum error correction \cite{HaPPY, FP2014, KC2018} that are directly analogous to the known quantum error correcting features of AdS/CFT \cite{ADH2015, DHW2016, CHPSSW2017}.

In this letter, we argue that tensor networks are not simply toy models for the AdS/CFT correspondence, but are in fact a precise, geometric description of static states in AdS/CFT over length scales greater than the string/Planck scales. Given a holographic CFT state with a static, semiclassical gravitational dual and a discretization of the corresponding bulk spacetime, we provide a general procedure for constructing a tensor network whose boundary state accurately reproduces the physics of the original CFT state and whose underlying geometry matches the chosen spacetime discretization.

For large-scale discretizations, our construction can be completed using known information-theoretic properties of holographic states. By using certain natural generalizations of the holographic entanglement of purification conjecture \cite{TU2017, NDHZS2017}, we can extend our construction to discretizations whose scale lies well below the AdS radius $\ell_{AdS}$ (so long as the discretization scale is still taken well above the string/Planck scales). If these conjectures hold up to stringy and quantum corrections, then our procedure can be used to construct a tensor network for any holographic CFT state for almost any discretization of the bulk. As a result, we claim that tensor networks are not mere ``toy models'': they describe the relationship between geometry and entanglement in \emph{actual} AdS/CFT!

In this letter, we first review the basic properties of tensor networks. We then construct a simple network for a bipartite discretization of vacuum $AdS_3$ and comment on how this procedure can be extended to generic ``tree'' discretizations formed by non-intersecting extremal surfaces. Finally, we introduce the holographic entanglement of purification conjecture, and show how it can be generalized to construct tensor networks with sub-AdS structure. The details of extending our construction to general spacetimes and general discretizations can be found in companion work \cite{BPSW2018}.

\section{Tensor Networks for Holographic States}

\subsection{Tensor Network Preliminaries} \label{sec:TN_review}

A multipartite quantum state $\ket{\psi} \in \mathcal{H} \equiv \mathcal{H}_{A_1} \otimes \dots \otimes \mathcal{H}_{A_n}$ may be thought of as a tensor with $n$ (abstract) up-indices $\psi^{A_1 \dots A_n}.$ When such a state is written as an outer product of other tensors, e.g.
\begin{equation} \label{eq:simple-network}
	\psi^{A_1 A_2} = S^{A_1 B} T^{A_2}{}_{B},
\end{equation}
then the contracted index $B$ may be thought of as a ``bond'' that combines two states
\begin{equation}
	\ket{S} \in \mathcal{H}_{A_1} \otimes \mathcal{H}_{B} \quad \text{and} \quad \ket{T} \in \mathcal{H}_{A_2} \otimes \mathcal{H}_{B}^*
\end{equation}
into a single state $\ket{\psi}.$ An outer-product representation of the form \eqref{eq:simple-network} for a multipartite tensor $\psi^{A_1 A_2}$ is called a \emph{tensor network} for $\ket{\psi}.$ Contracted Hilbert spaces like $\mathcal{H}_{B}$ are called \emph{bond spaces} in the network, with their dimensions referred to as \emph{bond dimensions}. When $\ket{\psi}$ has a restricted entanglement structure (as is the case for many physically interesting states), it is often possible to construct a tensor network for $\ket{\psi}$ that can be simulated efficiently on a classical computer \cite{Vidal2007, Vidal2008, VMC2008}.

\begin{figure}[h]
	\begin{center}
	\begin{tikzpicture}[->, >=latex, auto, tensor/.style={circle, draw, minimum width=3em}, label/.style={}]
		\node[tensor] (P) {$S$};
		\node[tensor] (Q) [right=2cm of P] {$T$};
		\node[label] (A1) [left=1cm of P] {$A_1$};
		\node[label] (A2) [right=1cm of Q] {$A_2$};

		\draw (P) edge[thick, bend left] node {$B$} (Q);
		\draw (P) edge[thick] node {} (A1);
		\draw (Q) edge[thick] node {} (A2);
	\end{tikzpicture}
	\caption{The graph corresponding to the tensor network expression in eq.\eqref{eq:simple-network}.}
	\label{fig:simple-network}
	\end{center}
\end{figure}
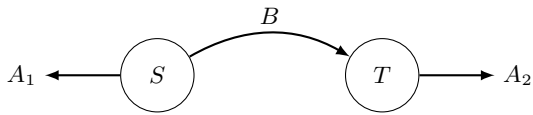

A tensor network can be represented as a graph where (i) each tensor in the expression corresponds to a vertex, (ii) each index corresponds to an edge with one end attached to that tensor, and (iii) contractions between indices are shown by linking the corresponding edges. In this sense, an arbitrary outer product representation of a multipartite quantum state defines a corresponding discrete geometry---a weighted graph with edge weights given by the bond dimensions of the corresponding index contractions. As an example, a graph for the tensor network given in \eqref{eq:simple-network} is sketched in Fig.\ref{fig:simple-network}. Note that in our convention, `up'-indices correspond to edges with outward-facing edges, while `down'-indices correspond to edges with inward-facing arrows.

\subsection{Smooth Entropies in Holographic States} \label{sec:tree}

Consider a state $\ket{\psi}$ in a large-$N$ CFT that is known to have a static, semiclassical gravitational dual. For any subregion $A$ of the CFT domain, the entanglement entropy of $\ket{\psi}$ between $A$ and its complement $A^c$ is given in a small-$G_N$ expansion by \cite{RT2006-1, FLM2013, EW2014, LM}
\begin{equation} \label{eq:RT}
	S = \frac{\operatorname{area}(\gamma)}{4 G_N} + O\left(1 \right),
\end{equation}
where $\gamma$, also known as the Ryu-Takayanagi (RT) surface, is the minimal codimension-2 bulk surface anchored on $\partial A$ and homologous to $A$.\footnote{In \eqref{eq:RT}, the CFT is regularized with an ultraviolet cutoff simultaneously with a corresponding radial cutoff in the bulk spacetime, so that both sides of the equation are finite \cite{RT2006-1}.}

The entanglement entropy of a subsystem, given by the von Neumann entropy $S = - \tr(\rho \log \rho)$ of the corresponding density matrix, is only one of many information-theoretic entropies that encode the entanglement spectrum of a state. More generally, we consider the \emph{R\'{e}nyi entropies}
\begin{equation}
	S_{\alpha}(\rho) = \frac{1}{1-\alpha} \log{\tr{\rho^{\alpha}}}. \quad (\alpha \geq 0)
\end{equation}
The R\'{e}nyi entropies are monotonically decreasing as a function of $\alpha$, and reproduce the von Neumann entropy in the limit $\alpha \rightarrow 1.$

In AdS/CFT, the R\'{e}nyi entropies of a subsystem in a holographic state can be computed as functions of the areas of surfaces in particular ``conical deficit'' geometries where multiple asymptotically-AdS geometries are glued together and allowed to backreact against one another \cite{Headrick2010, Dong2016}. Knowing the full spectrum of R\'{e}nyi entropies for a subsystem of a holographic state is sufficient to deduce its entanglement spectrum and hence the corresponding density matrix. These entropies are constrained somewhat by their monotonicity, since all R\'{e}nyi entropies must lie between the \emph{max-} and \emph{min-entropies}
\begin{eqnarray}
	S_{\text{max}}(\rho) & = \lim_{\alpha \rightarrow 0} S_{\alpha} = \log[\rank(\rho)], \\
	S_{\text{min}}(\rho) & = \lim_{\alpha \rightarrow \infty} S_{\alpha} = \log[\lambda_{\text{max}}^{-1}(\rho)].
\end{eqnarray}
(Here, $\lambda_{\text{max}}(\rho)$ denotes the largest eigenvalue of $\rho$.)

This constraint is rather weak in general, as there can be a large gap between the max- and min-entropies. For a holographic state, however, these max- and min-entropies can be tightly constrained without altering the physical properties of the state. Unpublished work due to Hayden, Swingle, and Walter \cite{HSW}, reproduced partially in \cite{BPSW2018}, shows that the \emph{smooth} min- and max-entropies of a subregion in a holographic state, defined by
\begin{eqnarray}
	S_{\text{max}}^{\varepsilon}(\rho) & = \min_{\onenorm{\sigma - \rho} < \varepsilon} S_{\text{max}}(\sigma), \\
	S_{\text{min}}^{\varepsilon}(\rho) & = \max_{\onenorm{\sigma - \rho} < \varepsilon} S_{\text{min}}(\sigma),
\end{eqnarray}
are given to leading order in $G_N$ by
\begin{eqnarray}
	S_{\text{max}}^{\varepsilon}(\rho) & = S + O\left(\sqrt{\frac{1}{G_N} \ln{\frac{1}{\varepsilon}}}\right) = S + O(\sqrt{S}), \label{eq:holog-max} \\
	S_{\text{min}}^{\varepsilon}(\rho) & = S - O\left(\sqrt{\frac{1}{G_N} \ln{\frac{1}{\varepsilon}}}\right) = S - O(\sqrt{S}), \label{eq:holog-min}
\end{eqnarray}
where $S$ is the von Neumann entropy.\footnote{These properties were first shown for certain special states in $1+1$-dimensional CFTs in \cite{CHLS2015}.} Here we use the trace norm $\onenorm{\sigma - \rho} = \tr(|\sigma - \rho|)$ as a metric on the space of density matrices. Equations \eqref{eq:holog-max} and \eqref{eq:holog-min} can be used to construct a \emph{nearby} state $\tilde{\rho}$ whose R\'{e}nyi entropies are tightly constrained around the von Neumann entropy. Since closeness of density matrices in the trace norm implies closeness of their expectation values on bounded operators, this `smoothed state' approximates the physical properties of the original state. The key point of this letter is that these approximate states can be expressed as tensor networks associated with asymptotically-AdS geometries.

\subsection{Holographic Networks from Entanglement Distillation}

Constraints on the smooth min- and max-entropies of holographic states allow us to perform \emph{entanglement distillation}, whereby a bipartite quantum state is approximated by a nearby state in which the entanglement between two regions is made manifest. In AdS/CFT, this `distilled' state is precisely a tensor network for the bipartite discretization of the corresponding spacetime. For simplicity, consider the $1+1$-dimensional vacuum state $\ket{\psi}$ whose holographic dual is vacuum $AdS_3,$ with reduced density matrix $\rho$ on a subregion $A$.

It was shown in \cite{BPSW2018} that the smooth min- and max-entropy properties of $\rho$ discussed above guarantee the existence of an approximate state $\ket{\Psi}$ of the form
\begin{equation} \label{eq:bipartite-TN-state}
	\ket{\Psi} = \sum_{n=0}^{e^{O(\sqrt{S})}} \sum_{m=0}^{e^{S - O(\sqrt{S})}} \sqrt{\lambda_{n}} \ket{n, m}_{A} \ket{n, m}_{A^c},
\end{equation}
where $\{n, m\}$ is an appropriately chosen division of the eigenvalues of the reduced state $\rho$ into $e^{O(\sqrt{S})}$ blocks of width $e^{S - O(\sqrt{S})}$. By calling $\ket{\Psi}$ an \emph{approximate state}, we mean that its overlap with the original CFT state $\ket{\psi}$ satisfies
\begin{equation} \label{eq:bipartite-error}
	\left|\braket{\Psi}{\psi}\right|^2 \geq 1 - \varepsilon - e^{-O(\sqrt{S})},
\end{equation}
and therefore that the expectation values of bounded operators in $\ket{\Psi}$ and $\ket{\psi}$ agree up to corrections of order $\sqrt{\varepsilon + e^{-O(\sqrt{S})}}.$

Since each $n$-block in eq.\eqref{eq:bipartite-TN-state} has a flat entanglement spectrum, we call $\ket{\Psi}$ a \emph{distilled state} for $\ket{\psi}$, as the entanglement between $A$ and $A^c$ in $\ket{\psi}$ has been `distilled' into maximally entangled states on Hilbert spaces of size $e^{O(S)}.$ To see these maximally entangled states explicitly, we introduce auxiliary Hilbert spaces $\mathcal{H}_{\gamma}$ and $\mathcal{H}_{f}$ with dimensions given by
\begin{equation} \label{eq:bond-dims}
	\dim\mathcal{H}_{f} = e^{O(\sqrt{S})} \quad \text{and} \quad \dim\mathcal{H}_{\gamma} = e^{S - O(\sqrt{S})}
\end{equation}
and define maps $\mathcal{H}_{f} \otimes \mathcal{H}_{\gamma} \rightarrow \mathcal{H}_{A}$ and $\bar{\mathcal{H}}_f \otimes \bar{\mathcal{H}}_{\gamma} \rightarrow \mathcal{H}_{A^c}$ by
\begin{eqnarray}
	V\ket{n}_{f} \ket{m}_{\gamma}
		& = \ket{n, m}_{A}, \\
	W\ket{\bar{n}}_{\bar{f}} \ket{\bar{m}}_{\bar{\gamma}}
		& = \ket{n, m}_{A^c}
\end{eqnarray}
for some arbitrarily chosen bases of the auxiliary Hilbert spaces and the corresponding bases in their complex conjugate Hilbert spaces.

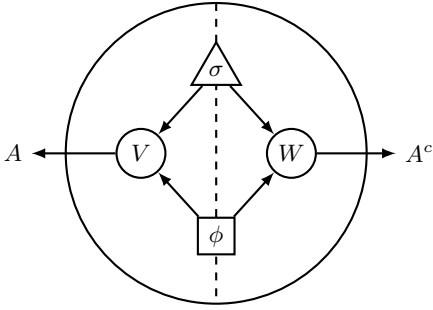
\begin{figure}
\begin{center}
	\begin{tikzpicture}[thick, >=latex, auto, vertex/.style={draw, shape=circle, fill=black, scale=0.7}, tensor/.style={circle, draw, minimum width=2em, inner sep=2pt},
				EPR/.style={regular polygon, regular polygon sides=4, fill=white, draw, inner sep=0.9pt},
				cobwebs/.style={regular polygon, fill=white, regular polygon sides=3, draw, inner sep=1pt}]
		% Boundary
		\draw (0, 0) circle (2cm);

		% RT Surface
		\draw [dashed] (0, 2) to (0, -2);

		% Tensors
		\node[tensor] (V) at (-1, 0) {$V$};
		\node[tensor] (W) at (1, 0) {$W$};
		\node[EPR] (phi) at (0, -1.1) {$\phi$};
		\node[cobwebs] (sigma) at (0, 1.1) {$\sigma$};

		% Boundary Labels
		\node[label] (A) at (-2.7, 0) {$A$};
		\node[label] (Ac) at (2.7, 0) {$A^c$};

		% Contractions
		\draw (phi) edge[->] node {} (V);
		\draw (phi) edge[->] node {} (W);
		\draw (sigma) edge[->] node {} (V);
		\draw (sigma) edge[->] node {} (W);
		\draw (V) edge[->] node {} (A);
		\draw (W) edge[->] node {} (Ac);
	\end{tikzpicture}
	\caption{A tensor network for a bipartite discretization of $AdS_3$ by a single Ryu-Takayanagi surface, drawn here as a dashed line in the bulk.}
	\label{fig:bipartite-network}
\end{center}
\end{figure}

The distilled state  $\ket{\Psi}$ may then be written as
\begin{equation} \label{eq:bipartite-distilled-state}
	\ket{\Psi}
		= (V \otimes W)
			(\ket{\phi} \otimes \ket{\sigma}),			
\end{equation}
with
\begin{eqnarray}
	\ket{\phi}
		&=&  \sum\limits_{m=0}^{e^{S - O(\sqrt{S})}} \ket{m \bar{m}}_{\gamma \bar{\gamma}}, \\
	\ket{\sigma}
		&=& \sum\limits_{n=0}^{e^{O(\sqrt{S})}} \sqrt{\lambda_{n}} \ket{n \bar{n}}_{f \bar{f}}.
\end{eqnarray}
This expression for $\ket{\Psi}$ is a tensor network of the form
\begin{equation} \label{eq:bipartite-TN}
	\Psi^{A A^c} = V^{A}{}_{f \gamma} W^{A^c}{}_{\bar{f} \bar{\gamma}} \phi^{\gamma \bar{\gamma}} \sigma^{f \bar{f}},
\end{equation}
with bond dimensions given by $|f| = |\bar{f}| = e^{O(\sqrt{S})}$ and $|\gamma| = |\bar{\gamma}| = e^{S - O(\sqrt{S})}.$ From the Ryu-Takayanagi formula \eqref{eq:RT}, we see that the bond dimensions are determined by the areas of a corresponding extremal surface in the bulk---the maximally entangled bonds in $\ket{\phi}$ have dimension $e^{\operatorname{area}/4 G_N}$ to leading order in $G_N$, while the submaximally entangled bonds in $\ket{\sigma}$ have subleading dimension in $G_N$ corresponding to the fact that in AdS/CFT, extremal surfaces in the bulk spacetime experience fluctuations in their areas due to, e.g., graviton effects. To make clear the correspondence between the spacetime geometry and the tensor network geometry, the graph for the network in eq.\eqref{eq:bipartite-TN} is sketched in Fig.\ref{fig:bipartite-network} over the corresponding discretization of vacuum $AdS_3$.

Since $\ket{\Psi}$ both approximates the physics of the original CFT state $\ket{\psi}$ in the sense of \eqref{eq:bipartite-error} and has a tensor network representation whose geometry matches certain known, discretized geometric features of the holographic dual to $\ket{\psi}$, we claim that the tensor network for $\ket{\Psi}$ given in \eqref{eq:bipartite-TN} constitutes an explicit tensor network for the AdS/CFT correspondence. The procedure detailed above can be generalized to any pure state in the AdS/CFT correspondence, not just the $AdS_3$ vacuum, and can be repeated inductively to construct an approximate tensor network for any discretization of the bulk by non-intersecting Ryu-Takayanagi surfaces. Because the underlying graphs of such networks contain no loops, they are referred to as \emph{tree tensor networks} \cite{SDV2006}. The details of this inductive procedure are reported in companion work \cite{BPSW2018}.

\section{Entanglement of Purification and Sub-AdS Structure}

Tree tensor networks alone cannot probe spacetime structure below the AdS scale, as the information contained in any bond is distributed nonlocally across an entire Ryu-Takayanagi surface. To construct tensor networks that probe sub-AdS structure, we propose a procedure for dividing bonds along RT surfaces based on the holographic entanglement of purification conjecture \cite{TU2017, NDHZS2017}.

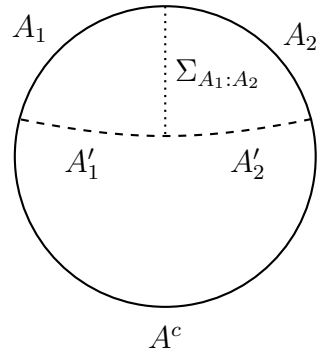
\begin{figure}[h]
\begin{center}
	\begin{tikzpicture}[thick, vertex/.style={draw, shape=circle, fill=black, scale=0.7}]
		% Boundary
		\draw (0, 0) circle (2cm);

		% RT Surface
		\draw [dashed] (1.95, 0.5) to[out=191.5, in=348.5] (-1.95, 0.5);

		% Entanglement Wedge Cross-section
		\draw [dotted] (0, 2) to (0, .3);

		% Boundary Labels
		\node at (-1.8, 1.7) (A1) {\large $A_1$};
		\node at (1.8, 1.6) (A2) {\large $A_2$};
		\node at (0, -2.4) (Ac) {\large $A^c$};

		% EW Label
		\node at (0.7, 1.1) (EW) {\large$\Sigma_{A_1 : A_2}$};

		% RT Labels
		\node at (-1.1, -0.1) (A1p) {\large$A_1'$};
		\node at (1.1, -0.1) (A2p) {\large$A_2'$};
	\end{tikzpicture}
	\caption{The entanglement wedge cross-section $\Sigma_{A_1 : A_2}$ for two neighboring boundary subregions $A_1$ and $A_2$ in vacuum $AdS_3.$}
	\label{fig:EP-EW}
\end{center}
\end{figure}

For a bipartite mixed state $\rho_{(A_1 A_2)}$, the \emph{entanglement of purification} between $A_1$ and $A_2$ is the minimal entanglement entropy between the two Hilbert spaces under a simultaneous purification \cite{THLD2002}, i.e.,
\begin{equation} \label{eq:eop}
	E_P(A_1:A_2)
		= \inf_{\ket{\Psi}_{A_1 A_1' A_2 A_2'}} S(A_1 A_1'),
\end{equation}
where $\ket{\Psi}_{A_1 A_1' A_2 A_2'}$ is any state that has $\rho_{(A_1 A_2)}$ as its reduced density matrix on $\mathcal{H}_{A_1} \otimes \mathcal{H}_{A_2}.$ The holographic entanglement of purification conjecture proposes that in AdS/CFT, this quantity corresponds to the area of the \emph{entanglement wedge cross-section}, which is the minimal bulk surface anchored to the Ryu-Takayanagi surface of $A_1 A_2$ that partitions the \emph{entanglement wedge} of $A_1 A_2$---i.e., the bulk region contained between $A_1 A_2$ and its RT surface---into a portion whose boundary contains all of $A_1$ and a disjoint portion whose boundary contains all of $A_2$. This surface is sketched in Fig.\ref{fig:EP-EW} in vacuum $AdS_3$ for a typical configuration of boundary regions $A_1$ and $A_2.$

\begin{figure}[h]
\begin{center}
	\subfloat[\label{fig:MEP-tree}]{
		\begin{tikzpicture}[thick, >=latex, scale=1,
					EPR/.style={regular polygon, regular polygon sides=4, fill=white, draw, inner sep=0.4pt},
					cobwebs/.style={regular polygon, fill=white, regular polygon sides=3, draw, inner sep=0.5pt},
					tensor/.style={circle, draw, fill=white, minimum width=1.5em, inner sep=2pt}]
		% Boundary
		\draw (1.95, 0.45) arc (13:167:2);

		% RT Surface
		\draw [dashed] (1.95, 0.45) to[out=191.5, in=348.5] (-1.95, 0.45);

		% Entanglement Wedge Cross-section
		\draw [dotted] (0, 2) to (0, .2);

		% Boundary Labels
		\node at (-1.9, 2.1) (A1) {\large $A_1$};
		\node at (1.9, 2.1) (A2) {\large $A_2$};

		% Girders and Cobwebs
		\node [EPR, inner sep=0.65pt] at (-0.7, 0.25) (phi1) {$\phi$};
		\node [cobwebs] at (-1.5, 0.3) (sigma1) {$\sigma$};
		\node [EPR] at (0, 0.7) (phi2) {$\phi$};
		\node [cobwebs] at (0, 1.3) (sigma2) {$\sigma$};
		\node [EPR, inner sep=0.6pt] at (0.7, 0.25) (phi3) {$\phi$};
		\node [cobwebs] at (1.5, 0.3) (sigma3) {$\sigma$};

		% Tensors
		\node [tensor] at (-1.8, -0.4) (T) {$T$};
		\node [tensor] at (-0.9, 1.1) (U) {$U$};
		\node [tensor] at (0.9, 1.1) (V) {$V$};
		\node [tensor] at (1.8, -0.4) (W) {$W$};

		% RT Labels
		\node at (-2.7, -0.7) (A1p) {\large $A_1'$};
		\node at (2.7, -0.7) (A2p) {\large $A_2'$};

		% External Edges
		\draw (T) edge[->] (A1p);
		\draw (U) edge[->] (A1);
		\draw (V) edge[->] (A2);
		\draw (W) edge[->] (A2p);

		% Internal Edges
		\draw (phi1) edge[->] (T);
		\draw (phi1) edge[->] (U);
		\draw (sigma1) edge[->] (T);
		\draw (sigma1) edge[->] (U);
		\draw (phi2) edge[->] (U);
		\draw (sigma2) edge[->] (U);
		\draw (phi2) edge[->] (V);
		\draw (sigma2) edge[->] (V);
		\draw (phi3) edge[->] (V);
		\draw (sigma3) edge[->] (V);
		\draw (phi3) edge[->] (W);
		\draw (sigma3) edge[->] (W);
		\end{tikzpicture}
	}
	\hspace{1cm}
	\subfloat[\label{fig:MEP-loop}]{
		\begin{tikzpicture}[thick, >=latex, scale=1,
					EPR/.style={regular polygon, regular polygon sides=4, fill=white, draw, inner sep=0.6pt},
					cobwebs/.style={regular polygon, fill=white, regular polygon sides=3, draw, inner sep=1pt},
					tensor/.style={circle, draw, fill=white, minimum width=1.5em, inner sep=2pt}]
		% Boundary
		\draw (0,0) circle (2cm);

		% RT Surface
		\draw [dashed] (1.95, 0.45) to[out=191.5, in=348.5] (-1.95, 0.45);

		% Entanglement Wedge Cross-section
		\draw [dotted] (0, 2) to (0, .2);

		% Boundary Labels
		\node at (-1.9, 2.1) (A1) {\large $A_1$};
		\node at (1.9, 2.1) (A2) {\large $A_2$};

		% Girders and Cobwebs
		\node [EPR, inner sep=0.65] at (-0.7, 0.25) (phi1) {$\phi$};
		\node [cobwebs] at (-1.5, 0.3) (sigma1) {$\sigma$};
		\node [EPR] at (0, 0.7) (phi2) {$\phi$};
		\node [cobwebs] at (0, 1.3) (sigma2) {$\sigma$};
		\node [EPR] at (0.7, 0.25) (phi3) {$\phi$};
		\node [cobwebs] at (1.5, 0.3) (sigma3) {$\sigma$};

		% Tensors
		\node [tensor] at (-0.9, 1.1) (U) {$U$};
		\node [tensor] at (0.9, 1.1) (V) {$V$};
		\node [tensor] at (0, -0.9) (X) {$Y$};

		% RT Labels
		\node at (-2.7, -0.7) (A1p) {\large $A_1'$};
		\node at (2.7, -0.7) (A2p) {\large $A_2'$};
		\node at (0, -2.6) (Ac) {\large $A^c$};

		% External Edges
		\draw (U) edge[->] (A1);
		\draw (V) edge[->] (A2);
		\draw (X) edge[->] (Ac);

		% Internal Edges
		\draw (phi1) edge[->] (X);
		\draw (phi1) edge[->] (U);
		\draw (sigma1) edge[->] (X);
		\draw (sigma1) edge[->] (U);
		\draw (phi2) edge[->] (U);
		\draw (sigma2) edge[->] (U);
		\draw (phi2) edge[->] (V);
		\draw (sigma2) edge[->] (V);
		\draw (phi3) edge[->] (V);
		\draw (sigma3) edge[->] (V);
		\draw (phi3) edge[->] (X);
		\draw (sigma3) edge[->] (X);

		\end{tikzpicture}
	}
	\caption{(a) A tensor network for the minimally entangled purification of two neighboring boundary regions in vacuum $AdS_3$.
			(b) The tensor network for the full boundary state formed by adding an isometry to the MEP.}
	\label{fig:MEP_networks}
\end{center}
\end{figure}
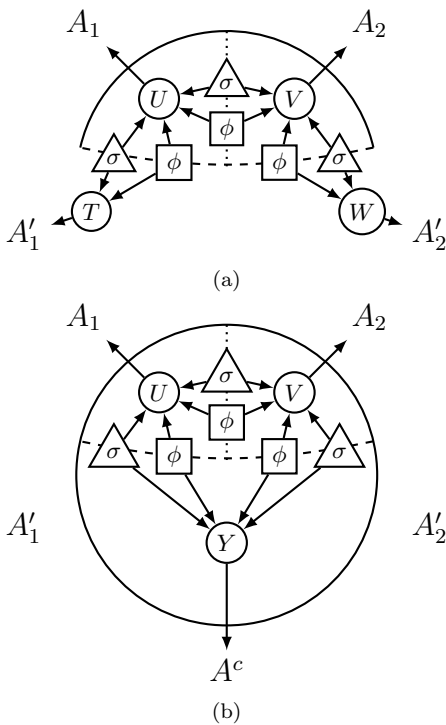

The holographic entanglement of purification conjecture associates the Hilbert space factors $A_1'$ and $A_2'$ in eq.\eqref{eq:eop} to the corresponding subregions of the RT surface, as sketched in Fig.\ref{fig:EP-EW}. More precisely, this conjecture invokes the surface-state correspondence \cite{MT2015} to claim that a state $\ket{\text{MEP}}$ (called the ``minimally entangled purification'') that saturates (or nearly saturates) the infimum in \eqref{eq:eop} is holographic for the entanglement wedge of $A_1 A_2$ in the sense that the entanglement entropies of its subregions can be computed by finding the areas of minimal surfaces in the entanglement wedge of $A_1 A_2$ in a manner analogous to the RT formula \eqref{eq:RT}. If $\ket{\text{MEP}}$ is indeed holographic for the bulk geometry of the entanglement wedge, then arguments given in the previous section suggest that it should be possible to build a tree tensor network for $\ket{\text{MEP}}$ whose geometry matches the region in Fig.\ref{fig:EP-EW} bounded by $A_1 A_2 A_1' A_2'.$  This is a tree tensor network corresponding to a discretization by three minimal surfaces ($A_1'$, $A_2',$ and $\Sigma_{A_1 : A_2}$), and is sketched in Fig.\ref{fig:MEP-tree} for neighboring boundary subregions in vacuum $AdS_3.$

Since $\ket{\text{MEP}}$ and the original global CFT state $\ket{\psi}$ are both purifications of $\rho_{(A_1 A_2)}$, they are related by an isometry operator $X$ that acts only on $A_1' A_2'$, i.e., $X \ket{\text{MEP}} = \ket{\psi}.$ Adding this isometry to the MEP tensor network sketched in Fig.\ref{fig:MEP-tree} (and defining $Y \equiv XTW$) yields a geometric tensor network for the boundary CFT state that corresponds to discretizing the bulk by the RT surface of $A_1 A_2$ and the entanglement wedge cross-section $\Sigma_{A_1 : A_2}.$ This tensor network manifestly represents the structure of \emph{divisions} of extremal surfaces in the bulk, and thus probes spacetime structure below the scale of the tree tensor networks described above.

By generalizing the holographic entanglement of purification conjecture to $n$ neighboring boundary regions, minimizing the quantity $S(A_1 A_1') + S(A_1 A_1' A_2 A_2') + \dots + S(A_1 A_1' \dots A_{n-1} A_{n-1}')$, one can construct tensor networks for discretizations of the bulk where a single Ryu-Takayanagi surface is split into $n$ subregions. The size of each subregion can be chosen to lie below the AdS scale and above the string/Planck scales, thus representing spacetime structure at scales below $\ell_{AdS}.$  Furthermore, if one assumes that the ``bottom half'' state composed of the tensor $Y$ in Fig.\ref{fig:MEP-loop} and neighboring edge states $\ket{\phi}$ and $\ket{\sigma}$ is holographic for the bottom half of the bulk spacetime in the same sense that $\ket{\text{MEP}}$ was assumed to be holographic for the top half of the bulk spacetime, then this procedure can be iterated to produce tensor networks for arbitrary grid-like discretizations of the bulk spacetime---even discretizations where the volume of space occupied by each tensor is well below $\ell_{AdS}^{d-1}$! The details of this procedure are reported in \cite{BPSW2018}.

\begin{acknowledgments}
We would like to acknowledge useful conversations with Ahmed Almheiri, Chris Akers, Raphael Bousso, Xi Dong, William Donnelly, Dan Harlow, Patrick Hayden, Ted Jacobson, Isaac Kim, Aitor Lewkowycz, Juan Maldacena, Don Marolf, Masamichi Miyaji, Ali Mollabashi, Xiao-Liang Qi, Dan Ranard, Eva Silverstein, Steve Shenker, Brian Swingle, Tadashi Takayanagi, and Guifre Vidal.  We especially thank Patrick Hayden, Brian Swingle, and Michael Walter for sharing with us their unpublished work on smooth min- and max-entropies in quantum field theory.  GP, JS, and AW were supported by the Simons Foundation (``It from Qubit''), AFOSR grant number FA9550-16-1-0082, and the Stanford Institute for Theoretical Physics.  AW was also supported in part by the John Templeton Foundation, Grant ID\# 60933, and would like to thank the Centro Atómico Bariloche, the California Institute of Technology, and the Kavli Institute for Theoretical Physics (supported in part by NSF grant PHY-1748958) for hospitality at key stages of this project.  NB is supported by the National Science Foundation under grant number 82248-13067-44-PHPXH and by the Department of Energy under grant number DE-SC0019380. He would also like to thank the SITP for hospitality while part of this work was completed.  The opinions expressed are those of the authors and do not necessarily reflect the views of funding agencies.
\end{acknowledgments}

\bibliography{holographic-networks-prl}

%apsrev4-2.bst 2018-12-27 (MD) hand-edited version of apsrev4-1.bst
\begin{thebibliography}{29}%
\makeatletter
\providecommand \@ifxundefined [1]{%
 \@ifx{#1\undefined}
}%
\providecommand \@ifnum [1]{%
 \ifnum #1\expandafter \@firstoftwo
 \else \expandafter \@secondoftwo
 \fi
}%
\providecommand \@ifx [1]{%
 \ifx #1\expandafter \@firstoftwo
 \else \expandafter \@secondoftwo
 \fi
}%
\providecommand \natexlab [1]{#1}%
\providecommand \enquote  [1]{``#1''}%
\providecommand \bibnamefont  [1]{#1}%
\providecommand \bibfnamefont [1]{#1}%
\providecommand \citenamefont [1]{#1}%
\providecommand \href@noop [0]{\@secondoftwo}%
\providecommand \href [0]{\begingroup \@sanitize@url \@href}%
\providecommand \@href[1]{\@@startlink{#1}\@@href}%
\providecommand \@@href[1]{\endgroup#1\@@endlink}%
\providecommand \@sanitize@url [0]{\catcode `\\12\catcode `\$12\catcode
  `\&12\catcode `\#12\catcode `\^12\catcode `\_12\catcode `\%12\relax}%
\providecommand \@@startlink[1]{}%
\providecommand \@@endlink[0]{}%
\providecommand \url  [0]{\begingroup\@sanitize@url \@url }%
\providecommand \@url [1]{\endgroup\@href {#1}{\urlprefix }}%
\providecommand \urlprefix  [0]{URL }%
\providecommand \Eprint [0]{\href }%
\providecommand \doibase [0]{https://doi.org/}%
\providecommand \selectlanguage [0]{\@gobble}%
\providecommand \bibinfo  [0]{\@secondoftwo}%
\providecommand \bibfield  [0]{\@secondoftwo}%
\providecommand \translation [1]{[#1]}%
\providecommand \BibitemOpen [0]{}%
\providecommand \bibitemStop [0]{}%
\providecommand \bibitemNoStop [0]{.\EOS\space}%
\providecommand \EOS [0]{\spacefactor3000\relax}%
\providecommand \BibitemShut  [1]{\csname bibitem#1\endcsname}%
\let\auto@bib@innerbib\@empty
%</preamble>
\bibitem [{\citenamefont {Maldacena}(1999)}]{Maldacena1999}%
  \BibitemOpen
  \bibfield  {author} {\bibinfo {author} {\bibfnamefont {J.}~\bibnamefont
  {Maldacena}},\ }\bibfield  {title} {\bibinfo {title} {The large-{N} limit of
  superconformal field theories and supergravity},\ }\href@noop {} {\bibfield
  {journal} {\bibinfo  {journal} {Int. J. Theor. Phys.}\ }\textbf {\bibinfo
  {volume} {38}},\ \bibinfo {pages} {1113} (\bibinfo {year} {1999})},\ \Eprint
  {https://arxiv.org/abs/hep-th/9711200} {arXiv:hep-th/9711200 [hep-th]}
  \BibitemShut {NoStop}%
\bibitem [{\citenamefont {{S. Ryu and T. Takayanagi}}(2006)}]{RT2006-1}%
  \BibitemOpen
  \bibfield  {author} {\bibinfo {author} {\bibnamefont {{S. Ryu and T.
  Takayanagi}}},\ }\bibfield  {title} {\bibinfo {title} {Holographic derivation
  of entanglement entropy from the anti-de-{S}itter space/conformal field
  theory correspondence},\ }\href@noop {} {\bibfield  {journal} {\bibinfo
  {journal} {Phys. Rev. Lett.}\ }\textbf {\bibinfo {volume} {96}},\ \bibinfo
  {pages} {181602} (\bibinfo {year} {2006})},\ \Eprint
  {https://arxiv.org/abs/hep-th/0603001} {arXiv:hep-th/0603001 [hep-th]}
  \BibitemShut {NoStop}%
\bibitem [{\citenamefont {{S. Ryu and T. Takayanagi}}()}]{RT2006-2}%
  \BibitemOpen
  \bibfield  {author} {\bibinfo {author} {\bibnamefont {{S. Ryu and T.
  Takayanagi}}},\ }\bibfield  {title} {\bibinfo {title} {Aspects of holographic
  entanglement entropy},\ }\href@noop {} {\bibfield  {journal} {\bibinfo
  {journal} {JHEP}\ }\textbf {\bibinfo {volume} {2006}}\bibinfo  {number} {
  (08)},\ \bibinfo {pages} {045}}\BibitemShut {NoStop}%
\bibitem [{\citenamefont {Hubeny}\ \emph {et~al.}(2007)\citenamefont {Hubeny},
  \citenamefont {Rangamani},\ and\ \citenamefont {Takayanagi}}]{HRT2007}%
  \BibitemOpen
\bibfield  {number} {  }\bibfield  {author} {\bibinfo {author} {\bibfnamefont
  {V.~E.}\ \bibnamefont {Hubeny}}, \bibinfo {author} {\bibfnamefont
  {M.}~\bibnamefont {Rangamani}}, and\ \bibinfo {author} {\bibfnamefont
  {T.}~\bibnamefont {Takayanagi}},\ }\bibfield  {title} {\bibinfo {title} {A
  covariant holographic entanglement entropy proposal},\ }\href@noop {}
  {\bibfield  {journal} {\bibinfo  {journal} {JHEP}\ }\textbf {\bibinfo
  {volume} {2007}}\bibfield  {number} {\bibinfo  {number} { (07)},\ \bibinfo
  {pages} {062}},\ }\Eprint {https://arxiv.org/abs/0705.0016} {arXiv:0705.0016
  [hep-th]} \BibitemShut {NoStop}%
\bibitem [{\citenamefont {Headrick}(2010)}]{Headrick2010}%
  \BibitemOpen
  \bibfield  {author} {\bibinfo {author} {\bibfnamefont {M.}~\bibnamefont
  {Headrick}},\ }\bibfield  {title} {\bibinfo {title} {Entanglement {R}\'{e}nyi
  entropies in holographic theories},\ }\href@noop {} {\bibfield  {journal}
  {\bibinfo  {journal} {Phys. Rev. D}\ }\textbf {\bibinfo {volume} {82}},\
  \bibinfo {pages} {126010} (\bibinfo {year} {2010})},\ \Eprint
  {https://arxiv.org/abs/1006.0047} {arXiv:1006.0047 [hep-th]} \BibitemShut
  {NoStop}%
\bibitem [{\citenamefont {Dong}(2016)}]{Dong2016}%
  \BibitemOpen
  \bibfield  {author} {\bibinfo {author} {\bibfnamefont {X.}~\bibnamefont
  {Dong}},\ }\bibfield  {title} {\bibinfo {title} {The gravity dual of
  {R}\'{e}nyi entropy},\ }\href@noop {} {\bibfield  {journal} {\bibinfo
  {journal} {Nature Communications}\ }\textbf {\bibinfo {volume} {7}},\
  \bibinfo {pages} {12472} (\bibinfo {year} {2016})},\ \Eprint
  {https://arxiv.org/abs/1601.06788} {arXiv:1601.06788 [hep-th]} \BibitemShut
  {NoStop}%
\bibitem [{\citenamefont {{A. Lewkowycz and J. Maldacena}}(2013)}]{LM}%
  \BibitemOpen
  \bibfield  {author} {\bibinfo {author} {\bibnamefont {{A. Lewkowycz and J.
  Maldacena}}},\ }\bibfield  {title} {\bibinfo {title} {Generalized
  gravitational entropy},\ }\href@noop {} {\bibfield  {journal} {\bibinfo
  {journal} {JHEP}\ }\textbf {\bibinfo {volume} {2013}}\bibfield  {number}
  {\bibinfo  {number} { (90)}},\ }\Eprint {https://arxiv.org/abs/1304.4926}
  {arXiv:1304.4926 [hep-th]} \BibitemShut {NoStop}%
\bibitem [{\citenamefont {Dong}\ \emph
  {et~al.}(2016{\natexlab{a}})\citenamefont {Dong}, \citenamefont {Lewkowycz},\
  and\ \citenamefont {Rangamani}}]{DLR}%
  \BibitemOpen
  \bibfield  {author} {\bibinfo {author} {\bibfnamefont {X.}~\bibnamefont
  {Dong}}, \bibinfo {author} {\bibfnamefont {A.}~\bibnamefont {Lewkowycz}},
  and\ \bibinfo {author} {\bibfnamefont {M.}~\bibnamefont {Rangamani}},\
  }\bibfield  {title} {\bibinfo {title} {Deriving covariant holographic
  entanglement},\ }\href@noop {} {\bibfield  {journal} {\bibinfo  {journal}
  {JHEP}\ }\textbf {\bibinfo {volume} {2016}}\bibfield  {number} {\bibinfo
  {number} { (28)}},\ }\Eprint {https://arxiv.org/abs/1607.07506}
  {arXiv:1607.07506 [hep-th]} \BibitemShut {NoStop}%
\bibitem [{\citenamefont {Swingle}(2012{\natexlab{a}})}]{Swingle2012-1}%
  \BibitemOpen
  \bibfield  {author} {\bibinfo {author} {\bibfnamefont {B.}~\bibnamefont
  {Swingle}},\ }\bibfield  {title} {\bibinfo {title} {Entanglement
  renormalization and holography},\ }\href@noop {} {\bibfield  {journal}
  {\bibinfo  {journal} {Phys. Rev. D}\ }\textbf {\bibinfo {volume} {86}},\
  \bibinfo {pages} {065007} (\bibinfo {year} {2012}{\natexlab{a}})},\ \Eprint
  {https://arxiv.org/abs/0905.1317} {arXiv:0905.1317 [cond-mat.str-el]}
  \BibitemShut {NoStop}%
\bibitem [{\citenamefont {Swingle}(2012{\natexlab{b}})}]{Swingle2012-2}%
  \BibitemOpen
  \bibfield  {author} {\bibinfo {author} {\bibfnamefont {B.}~\bibnamefont
  {Swingle}},\ }\bibfield  {title} {\bibinfo {title} {Constructing holographic
  spacetimes using entanglement renormalization},\ }\href@noop {} {\  (\bibinfo
  {year} {2012}{\natexlab{b}})},\ \Eprint {https://arxiv.org/abs/1209.3304}
  {arXiv:1209.3304 [hep-th]} \BibitemShut {NoStop}%
\bibitem [{\citenamefont {Pastawski}\ \emph {et~al.}(2015)\citenamefont
  {Pastawski}, \citenamefont {Yoshida}, \citenamefont {Harlow},\ and\
  \citenamefont {Preskill}}]{HaPPY}%
  \BibitemOpen
  \bibfield  {author} {\bibinfo {author} {\bibfnamefont {F.}~\bibnamefont
  {Pastawski}}, \bibinfo {author} {\bibfnamefont {B.}~\bibnamefont {Yoshida}},
  \bibinfo {author} {\bibfnamefont {D.}~\bibnamefont {Harlow}}, and\ \bibinfo
  {author} {\bibfnamefont {J.}~\bibnamefont {Preskill}},\ }\bibfield  {title}
  {\bibinfo {title} {Holographic quantum error-correcting codes: Toy models for
  the bulk/boundary correspondence},\ }\href@noop {} {\bibfield  {journal}
  {\bibinfo  {journal} {JHEP}\ }\textbf {\bibinfo {volume} {2015}}\bibfield
  {number} {\bibinfo  {number} { (6)},\ \bibinfo {pages} {149}},\ }\Eprint
  {https://arxiv.org/abs/1503.06237} {arXiv:1503.06237 [hep-th]} \BibitemShut
  {NoStop}%
\bibitem [{\citenamefont {{A. J. Ferris and D. Poulin}}(2014)}]{FP2014}%
  \BibitemOpen
  \bibfield  {author} {\bibinfo {author} {\bibnamefont {{A. J. Ferris and D.
  Poulin}}},\ }\bibfield  {title} {\bibinfo {title} {Tensor networks and
  quantum error correction},\ }\href@noop {} {\bibfield  {journal} {\bibinfo
  {journal} {Phys. Rev. Lett.}\ }\textbf {\bibinfo {volume} {113}},\ \bibinfo
  {pages} {030501} (\bibinfo {year} {2014})},\ \Eprint
  {https://arxiv.org/abs/1312.4578} {arXiv:1312.4578 [quant-ph]} \BibitemShut
  {NoStop}%
\bibitem [{\citenamefont {{T. Kohler and T. Cubitt}}()}]{KC2018}%
  \BibitemOpen
  \bibfield  {author} {\bibinfo {author} {\bibnamefont {{T. Kohler and T.
  Cubitt}}},\ }\bibfield  {title} {\bibinfo {title} {Complete toy models of
  holographic duality},\ }\href@noop {} {\ }\Eprint
  {https://arxiv.org/abs/1810.08992} {arXiv:1810.08992 [hep-th]} \BibitemShut
  {NoStop}%
\bibitem [{\citenamefont {Almheiri}\ \emph {et~al.}(2015)\citenamefont
  {Almheiri}, \citenamefont {Dong},\ and\ \citenamefont {Harlow}}]{ADH2015}%
  \BibitemOpen
  \bibfield  {author} {\bibinfo {author} {\bibfnamefont {A.}~\bibnamefont
  {Almheiri}}, \bibinfo {author} {\bibfnamefont {X.}~\bibnamefont {Dong}}, and\
  \bibinfo {author} {\bibfnamefont {D.}~\bibnamefont {Harlow}},\ }\bibfield
  {title} {\bibinfo {title} {Bulk locality and quantum error correction in
  {AdS/CFT}},\ }\href@noop {} {\bibfield  {journal} {\bibinfo  {journal}
  {JHEP}\ }\textbf {\bibinfo {volume} {2015}}\bibfield  {number} {\bibinfo
  {number} { (4)},\ \bibinfo {pages} {163}},\ }\Eprint
  {https://arxiv.org/abs/1411.7041} {arXiv:1411.7041 [hep-th]} \BibitemShut
  {NoStop}%
\bibitem [{\citenamefont {Dong}\ \emph
  {et~al.}(2016{\natexlab{b}})\citenamefont {Dong}, \citenamefont {Harlow},\
  and\ \citenamefont {Wall}}]{DHW2016}%
  \BibitemOpen
  \bibfield  {author} {\bibinfo {author} {\bibfnamefont {X.}~\bibnamefont
  {Dong}}, \bibinfo {author} {\bibfnamefont {D.}~\bibnamefont {Harlow}}, and\
  \bibinfo {author} {\bibfnamefont {A.~C.}\ \bibnamefont {Wall}},\ }\bibfield
  {title} {\bibinfo {title} {Reconstruction of bulk operators within the
  entanglement wedge in gauge-gravity duality},\ }\href@noop {} {\bibfield
  {journal} {\bibinfo  {journal} {Phys. Rev. Lett.}\ }\textbf {\bibinfo
  {volume} {117}},\ \bibinfo {pages} {021601} (\bibinfo {year}
  {2016}{\natexlab{b}})},\ \Eprint {https://arxiv.org/abs/1601.05416}
  {arXiv:1601.05416 [hep-th]} \BibitemShut {NoStop}%
\bibitem [{\citenamefont {Cotler}\ \emph {et~al.}(2017)\citenamefont {Cotler},
  \citenamefont {Hayden}, \citenamefont {Penington}, \citenamefont {Salton},
  \citenamefont {Swingle},\ and\ \citenamefont {Walter}}]{CHPSSW2017}%
  \BibitemOpen
  \bibfield  {author} {\bibinfo {author} {\bibfnamefont {J.}~\bibnamefont
  {Cotler}}, \bibinfo {author} {\bibfnamefont {P.}~\bibnamefont {Hayden}},
  \bibinfo {author} {\bibfnamefont {G.}~\bibnamefont {Penington}}, \bibinfo
  {author} {\bibfnamefont {G.}~\bibnamefont {Salton}}, \bibinfo {author}
  {\bibfnamefont {B.}~\bibnamefont {Swingle}}, and\ \bibinfo {author}
  {\bibfnamefont {M.}~\bibnamefont {Walter}},\ }\bibfield  {title} {\bibinfo
  {title} {Entanglement wedge reconstruction via universal recovery channels},\
  }\href@noop {} {\  (\bibinfo {year} {2017})},\ \Eprint
  {https://arxiv.org/abs/1704.05839} {arXiv:1704.05839 [hep-th]} \BibitemShut
  {NoStop}%
\bibitem [{\citenamefont {{T. Takayanagi and K. Umemoto}}(2017)}]{TU2017}%
  \BibitemOpen
  \bibfield  {author} {\bibinfo {author} {\bibnamefont {{T. Takayanagi and K.
  Umemoto}}},\ }\bibfield  {title} {\bibinfo {title} {Holographic entanglement
  of purification},\ }\href@noop {} {\  (\bibinfo {year} {2017})},\ \Eprint
  {https://arxiv.org/abs/1708.09393} {arXiv:1708.09393 [hep-th]} \BibitemShut
  {NoStop}%
\bibitem [{\citenamefont {Nguyen}\ \emph {et~al.}(2017)\citenamefont {Nguyen},
  \citenamefont {Devakul}, \citenamefont {Halbasch}, \citenamefont {Zalatel},\
  and\ \citenamefont {Swingle}}]{NDHZS2017}%
  \BibitemOpen
  \bibfield  {author} {\bibinfo {author} {\bibfnamefont {P.}~\bibnamefont
  {Nguyen}}, \bibinfo {author} {\bibfnamefont {T.}~\bibnamefont {Devakul}},
  \bibinfo {author} {\bibfnamefont {M.~G.}\ \bibnamefont {Halbasch}}, \bibinfo
  {author} {\bibfnamefont {M.~P.}\ \bibnamefont {Zalatel}}, and\ \bibinfo
  {author} {\bibfnamefont {B.}~\bibnamefont {Swingle}},\ }\bibfield  {title}
  {\bibinfo {title} {Entanglement of purification: from spin chains to
  holography},\ }\href@noop {} {\bibfield  {journal} {\bibinfo  {journal}
  {JHEP}\ }\textbf {\bibinfo {volume} {2018}}\bibfield  {number} {\bibinfo
  {number} { (98)}},\ }\Eprint {https://arxiv.org/abs/1709.07424}
  {arXiv:1709.07424 [hep-th]} \BibitemShut {NoStop}%
\bibitem [{\citenamefont {Bao}\ \emph {et~al.}(2018)\citenamefont {Bao},
  \citenamefont {Penington}, \citenamefont {Sorce},\ and\ \citenamefont
  {Wall}}]{BPSW2018}%
  \BibitemOpen
  \bibfield  {author} {\bibinfo {author} {\bibfnamefont {N.}~\bibnamefont
  {Bao}}, \bibinfo {author} {\bibfnamefont {G.}~\bibnamefont {Penington}},
  \bibinfo {author} {\bibfnamefont {J.}~\bibnamefont {Sorce}}, and\ \bibinfo
  {author} {\bibfnamefont {A.~C.}\ \bibnamefont {Wall}},\ }\bibfield  {title}
  {\bibinfo {title} {Beyond toy models: Distilling tensor networks in full
  {AdS/CFT}},\ }\href@noop {} {\  (\bibinfo {year} {2018})},\ \Eprint
  {https://arxiv.org/abs/1812.01171} {arXiv:1812.01171 [hep-th]} \BibitemShut
  {NoStop}%
\bibitem [{\citenamefont {Vidal}(2007)}]{Vidal2007}%
  \BibitemOpen
  \bibfield  {author} {\bibinfo {author} {\bibfnamefont {G.}~\bibnamefont
  {Vidal}},\ }\bibfield  {title} {\bibinfo {title} {Entanglement
  renormalization},\ }\href@noop {} {\bibfield  {journal} {\bibinfo  {journal}
  {Phys. Rev. Lett.}\ }\textbf {\bibinfo {volume} {99}},\ \bibinfo {pages}
  {220405} (\bibinfo {year} {2007})},\ \Eprint
  {https://arxiv.org/abs/cond-mat/0512165} {arXiv:cond-mat/0512165 [cond-mat]}
  \BibitemShut {NoStop}%
\bibitem [{\citenamefont {Vidal}(2008)}]{Vidal2008}%
  \BibitemOpen
  \bibfield  {author} {\bibinfo {author} {\bibfnamefont {G.}~\bibnamefont
  {Vidal}},\ }\bibfield  {title} {\bibinfo {title} {Class of quantum many-body
  states that can be efficiently simulated},\ }\href@noop {} {\bibfield
  {journal} {\bibinfo  {journal} {Phys. Rev. Lett.}\ }\textbf {\bibinfo
  {volume} {101}},\ \bibinfo {pages} {110501} (\bibinfo {year} {2008})},\
  \Eprint {https://arxiv.org/abs/0610099} {arXiv:0610099 [quant-ph]}
  \BibitemShut {NoStop}%
\bibitem [{\citenamefont {Verstraete}\ \emph {et~al.}(2008)\citenamefont
  {Verstraete}, \citenamefont {Murg},\ and\ \citenamefont {Cirac}}]{VMC2008}%
  \BibitemOpen
  \bibfield  {author} {\bibinfo {author} {\bibfnamefont {F.}~\bibnamefont
  {Verstraete}}, \bibinfo {author} {\bibfnamefont {V.}~\bibnamefont {Murg}},
  and\ \bibinfo {author} {\bibfnamefont {J.~I.}\ \bibnamefont {Cirac}},\
  }\bibfield  {title} {\bibinfo {title} {Matrix product states, projected
  entangled pair states, and variational renormalization group methods for
  quantum spin systems},\ }\href@noop {} {\bibfield  {journal} {\bibinfo
  {journal} {Adv. in Phys.}\ }\textbf {\bibinfo {volume} {57}},\ \bibinfo
  {pages} {143} (\bibinfo {year} {2008})},\ \Eprint
  {https://arxiv.org/abs/0907.2796} {arXiv:0907.2796 [quant-ph]} \BibitemShut
  {NoStop}%
\bibitem [{\citenamefont {Faulkner}\ \emph {et~al.}(2013)\citenamefont
  {Faulkner}, \citenamefont {Lewkowycz},\ and\ \citenamefont
  {Maldacena}}]{FLM2013}%
  \BibitemOpen
  \bibfield  {author} {\bibinfo {author} {\bibfnamefont {T.}~\bibnamefont
  {Faulkner}}, \bibinfo {author} {\bibfnamefont {A.}~\bibnamefont {Lewkowycz}},
  and\ \bibinfo {author} {\bibfnamefont {J.}~\bibnamefont {Maldacena}},\
  }\bibfield  {title} {\bibinfo {title} {Quantum corrections to holographic
  entanglement entropy},\ }\href@noop {} {\bibfield  {journal} {\bibinfo
  {journal} {JHEP}\ }\textbf {\bibinfo {volume} {2013}}\bibfield  {number}
  {\bibinfo  {number} { (74)}},\ }\Eprint {https://arxiv.org/abs/1307.2892}
  {arXiv:1307.2892 [hep-th]} \BibitemShut {NoStop}%
\bibitem [{\citenamefont {{N. Engelhardt and A. C. Wall}}(2014)}]{EW2014}%
  \BibitemOpen
  \bibfield  {author} {\bibinfo {author} {\bibnamefont {{N. Engelhardt and A.
  C. Wall}}},\ }\bibfield  {title} {\bibinfo {title} {Quantum extremal
  surfaces: Holographic entanglement entropy beyond the classical regime},\
  }\href@noop {} {\bibfield  {journal} {\bibinfo  {journal} {JHEP}\ }\textbf
  {\bibinfo {volume} {2015}}\bibfield  {number} {\bibinfo  {number} { (73)}},\
  }\Eprint {https://arxiv.org/abs/1408.3203} {arXiv:1408.3203 [hep-th]}
  \BibitemShut {NoStop}%
\bibitem [{\citenamefont {Hayden}\ \emph {et~al.}()\citenamefont {Hayden},
  \citenamefont {Swingle},\ and\ \citenamefont {Walter}}]{HSW}%
  \BibitemOpen
  \bibfield  {author} {\bibinfo {author} {\bibfnamefont {P.}~\bibnamefont
  {Hayden}}, \bibinfo {author} {\bibfnamefont {B.}~\bibnamefont {Swingle}},
  and\ \bibinfo {author} {\bibfnamefont {M.}~\bibnamefont {Walter}},\
  }\bibfield  {title} {\bibinfo {title} {One-shot information theory in quantum
  field theory},\ }\href@noop {} {\bibinfo  {journal} {unpublished}\
  }\BibitemShut {NoStop}%
\bibitem [{\citenamefont {Czech}\ \emph {et~al.}(2015)\citenamefont {Czech},
  \citenamefont {Hayden}, \citenamefont {Lashkar},\ and\ \citenamefont
  {Swingle}}]{CHLS2015}%
  \BibitemOpen
\bibfield  {journal} {  }\bibfield  {author} {\bibinfo {author} {\bibfnamefont
  {B.}~\bibnamefont {Czech}}, \bibinfo {author} {\bibfnamefont
  {P.}~\bibnamefont {Hayden}}, \bibinfo {author} {\bibfnamefont
  {N.}~\bibnamefont {Lashkar}}, and\ \bibinfo {author} {\bibfnamefont
  {B.}~\bibnamefont {Swingle}},\ }\bibfield  {title} {\bibinfo {title} {The
  information theoretic interpretation of the length of a curve},\ }\href@noop
  {} {\bibfield  {journal} {\bibinfo  {journal} {JHEP}\ }\textbf {\bibinfo
  {volume} {2015:157}}},\ \Eprint {https://arxiv.org/abs/1410.1540}
  {arXiv:1410.1540 [hep-th]} \BibitemShut {NoStop}%
\bibitem [{\citenamefont {Shi}\ \emph {et~al.}(2006)\citenamefont {Shi},
  \citenamefont {Duan},\ and\ \citenamefont {Vidal}}]{SDV2006}%
  \BibitemOpen
  \bibfield  {author} {\bibinfo {author} {\bibfnamefont {Y.}~\bibnamefont
  {Shi}}, \bibinfo {author} {\bibfnamefont {L.}~\bibnamefont {Duan}}, and\
  \bibinfo {author} {\bibfnamefont {G.}~\bibnamefont {Vidal}},\ }\bibfield
  {title} {\bibinfo {title} {Classical simulation of quantum many-body systems
  with a tree tensor network},\ }\href@noop {} {\bibfield  {journal} {\bibinfo
  {journal} {Phys. Rev. A}\ }\textbf {\bibinfo {volume} {74}},\ \bibinfo
  {pages} {022320} (\bibinfo {year} {2006})},\ \Eprint
  {https://arxiv.org/abs/quant-ph/0511070} {arXiv:quant-ph/0511070 [quant-ph]}
  \BibitemShut {NoStop}%
\bibitem [{\citenamefont {Terhal}\ \emph {et~al.}(2002)\citenamefont {Terhal},
  \citenamefont {Horodecki}, \citenamefont {Leung},\ and\ \citenamefont
  {DiVincenzo}}]{THLD2002}%
  \BibitemOpen
  \bibfield  {author} {\bibinfo {author} {\bibfnamefont {B.~M.}\ \bibnamefont
  {Terhal}}, \bibinfo {author} {\bibfnamefont {M.}~\bibnamefont {Horodecki}},
  \bibinfo {author} {\bibfnamefont {D.~W.}\ \bibnamefont {Leung}}, and\
  \bibinfo {author} {\bibfnamefont {D.~P.}\ \bibnamefont {DiVincenzo}},\
  }\bibfield  {title} {\bibinfo {title} {The entanglement of purification},\
  }\href@noop {} {\bibfield  {journal} {\bibinfo  {journal} {J. Math. Phys}\
  }\textbf {\bibinfo {volume} {43}},\ \bibinfo {pages} {4286} (\bibinfo {year}
  {2002})},\ \Eprint {https://arxiv.org/abs/quant-ph/0202044}
  {arXiv:quant-ph/0202044 [quant-ph]} \BibitemShut {NoStop}%
\bibitem [{\citenamefont {{M. Miyaji and T. Takayanagi}}(2015)}]{MT2015}%
  \BibitemOpen
  \bibfield  {author} {\bibinfo {author} {\bibnamefont {{M. Miyaji and T.
  Takayanagi}}},\ }\bibfield  {title} {\bibinfo {title} {Surface/state
  correspondence as a generalized holography},\ }\href@noop {} {\bibfield
  {journal} {\bibinfo  {journal} {Prog. Theor. Exp. Phys.}\ } (\bibinfo {year}
  {2015})},\ \Eprint {https://arxiv.org/abs/1503.03542} {arXiv:1503.03542
  [hep-th]} \BibitemShut {NoStop}%
\end{thebibliography}%

\end{document}